\documentclass[journal,article,accept,moreauthors,pdftex,12pt,a4paper]{mdpi} 
 \lastpage{378} \doinum{10.3390/e13020367}
\pubvolume{13} \pubyear{2011} \history{Received: 23 December 2010;
in revised form: 19 January 2011 / Accepted: 20 January 2011 /
Published: 31 January 2011}
\usepackage{hyphenat}
\usepackage[sort&compress]{natbib}
\usepackage{float}
\usepackage{amssymb,amsmath,graphicx}

\Title{Lagrange Equations Coupled to a Thermal Equation: Mechanics
as Consequence of Thermodynamics}

\Author{Christian Gruber $^{1,\star}$ and Sylvain D. Brechet $^{2}$}

\address{%
$^{1}$ Institute of Theoretical Physics, Station 3, Ecole
Polytechnique F\'ed\'erale de Lausanne,
CH-1015 Lausanne, Switzerland\\
$^{2}$ Institute of Condensed Matter Physics, Station 3, Ecole
Polytechnique F\'ed\'erale de Lausanne,
CH-1015 Lausanne, Switzerland; E-Mail: sylvain.brechet@epfl.ch\\
}

\corres{E-Mail: christian.gruber@epfl.ch.}

\abstract{Following the analytic approach to thermodynamics
developed by St\"uckelberg, we study the evolution equations of a
closed thermodynamic system consisting of point particles in a
fluid. We obtain a system of coupled differential equations
describing the mechanical and the thermal evolution of the system.
The coupling between these evolution equations is due to the action
of a viscous friction term. Finally, we apply our coupled evolution
equations to study the thermodynamics of an isolated system
consisting of identical point particles interacting through a
harmonic potential. }


\begin{document}

\section{Introduction}

The history and development of modern thermodynamics started during
the industrialisation period and was closely related to the need to
improve the efficiency of steam engines. Since the foundational work
of Carnot~\cite{Carnot:1824} in 1824, until 1960, thermodynamics has
been a phenomenological theory mainly restricted to the description
of either equilibrium states or transformations relating one
equilibrium state to another~\cite{Ferrari:2010}. In essence,
thermodynamics was essentially reduced to thermostatics or
quasi-thermostatics.

In 1960, St\"uckelberg~\cite{Stueckelberg:1974} reformulated
thermodynamics in an axiomatic way as a truly dynamic and
phenomenological theory describing the evolution of a thermodynamic
system by a set of first-order differential equations. By doing so,
he actually extended the existing theory of equilibrium states, or
thermostatic, in order to obtain a genuine thermodynamic theory,
{\em i.e.}, a theory describing the evolution and the approach to
equilibrium of thermodynamic systems. In his axiomatic approach, he
introduced, two state functions, the ``energy'' and the ``entropy''
obeying the two fundamental laws of thermodynamics. He then derived
from these two laws the equations for the time evolution of the
system.

In his words, the state of a thermodynamic system is defined by a
set of geometric variables and a set of thermal variables. The
simplest system, which he called ``system element'', is a system
where one thermal variable and a set of geometric variables are
sufficient to define entirely the state. The second law of
thermodynamics requires the existence of a state function entropy
for every system. Thus, the state of a system element is defined by
the entropy and a set of geometric variables. We note that the
notion of system element is analogous to the concept of point
particle in mechanics. It is the building block required to develop
the theory of general systems where several thermal variables are
necessary to define the state.

In the present work, we essentially follow the original approach of
St\"uckelberg~\cite{Stueckelberg:1974} and apply his formalism to a
closed thermodynamic system consisting of $N$ point particles moving
in a fluid. The system is thus defined by the particles and the
fluid, but we assume no a priori knowledge of the fluid. It is a
phenomenological approach where the theory is built upon the
macroscopic description of the particles (which are the only
observable objects). It is thus observed that the macroscopical
mechanical variables (generalised coordinates and velocities) are
not sufficient to obtain the time evolution of the system. Following
the thermodynamic approach, we assume the simplest possible case
where the system can be described by introducing only one more
non-mechanical, or thermodynamical variable, the entropy. Applying
the first and second law, we obtain a system of coupled differential
equations describing the thermodynamic evolution of the system. In
that system of differential equations, the coupling between the
resulting Lagrange equations and the thermal equation is due to the
viscous friction terms depending on the state variable.

In this phenomenological approach, the theory tells us what are the
quantities that should be obtained from experiment, in particular in
order to obtain typical thermodynamical properties such as the
specific heat. Since we have introduced a model with only one
thermal variable (the total entropy), it is not possible to derive
transport properties (viscosity or thermal conductivity). To obtain
such a transport theory, we would need a continuum description of
the fluid (density field, entropy field, ...).

The structure of this publication is as follows. In Section~\ref{2},
we briefly recall St\"uckelberg's axiomatic formulation of the first
and second thermodynamic laws. Section~\ref{3} is devoted to
establishing the equations of evolution of a system of point
particles based on the first law. In Section~\ref{4}, we extend the
thermodynamics of a system of point particles to include the second
law. Section~\ref{5} shows how thermodynamics relates to mechanics
and in particular to the Lagrange equations. Finally, in
Section~\ref{6}, we apply our formalism to study the thermodynamics
of an isolated system consisting of identical point particles
interacting through a harmonic potential, which is the simplest
phenomenological model of a~solid.

\section{Axiomatic Formulation of the First and Second Laws}
\label{2}

\subsection{First Law}
For every system $\Sigma$, there exists an extensive, scalar state
function $E$, called energy. If the system is isolated, the energy
is constant, {\em i.e.}, $E$ is a conserved observable. If the
system is not isolated, then
\begin{equation}
\label{first} \frac{dE}{dt} = P_W^{\,\text{ext}}(t) +
P_Q^{\,\text{ext}}(t) + P_{\text{chem}}^{\,\text{ext}}(t)\ ,
\end{equation}
where $P_W^{\,\text{ext}}(t)$ is the power due to the external
forces acting on the mechanical variables of the system,
$P_Q^{\,\text{ext}}(t)$ is the power due to the heat transfer and
$P_{\text{chem}}^{\,\text{ext}}(t)$ is the power due to the matter
transfer between the system and the exterior
(see~~\cite{Stueckelberg:1974} p.~26 and~\cite{Fuchs:1996} p.~221).

The system is said to be ``closed'' if there is no exchange of
matter, {\em i.e.}, $P_{\text{chem}}^{\,\text{ext}}(t)=0$~; it is
said ``adiabatically closed'' if it is closed and there is no heat
exchange, {\em i.e.}, $P_{\text{chem}}^{\,\text{ext}}(t)=0$ and
$P_Q^{\,\text{ext}}(t)=0$. It is said isolated if it is
adiabatically closed and there is no mechanical power exchange, {\em
i.e.}, $P_{\text{chem}}^{\,\text{ext}}(t)=0$,
$P_Q^{\,\text{ext}}(t)=0$ and $P_W^{\,\text{ext}}(t)=0$, in which
case the energy $E$ is a constant.

\subsection{Second Law}

For every system $\Sigma$, there exists an extensive, scalar state
function $S$, called entropy, which obeys the following two
conditions~(see~\cite{Stueckelberg:1974} p.~23)~:
\begin{enumerate}
\item[(a)] {\it Evolution Part~:}\\
If the system is adiabatically closed, the entropy $S$ is a
non-decreasing function with respect to time, {\em i.e.},
\begin{equation}
\label{entropy_prod} \frac{dS}{dt} = I(t) \geqslant 0\ ,
\end{equation}
where $I(t)$ is the entropy production rate of the system accounting
for the irreversibility of internal thermal processes.
\item[(b)] {\it Equilibrium Part~:}\\
If the system is isolated, as time tends to infinity ({\em i.e.},
$t\rightarrow +\infty$) the entropy tends towards a finite local
maximum~\cite{Gruber:1999}, compatible with the constraints
(internal walls and isolation conditions), {\em i.e.},
\begin{equation}
\label{entropy_max} \underset{t\rightarrow +\infty}{\lim} S(t) =
\max_{\substack{\rho}\ \text{compatible}} S[\rho]\ ,
\end{equation}
where $\rho$ compatible denotes a thermodynamic state compatible
with the constraints.
\end{enumerate}


\section{System of Point Particles: First Law}
\label{3}

\begin{figure}[H]
\centering
\includegraphics[width=0.62\textwidth]{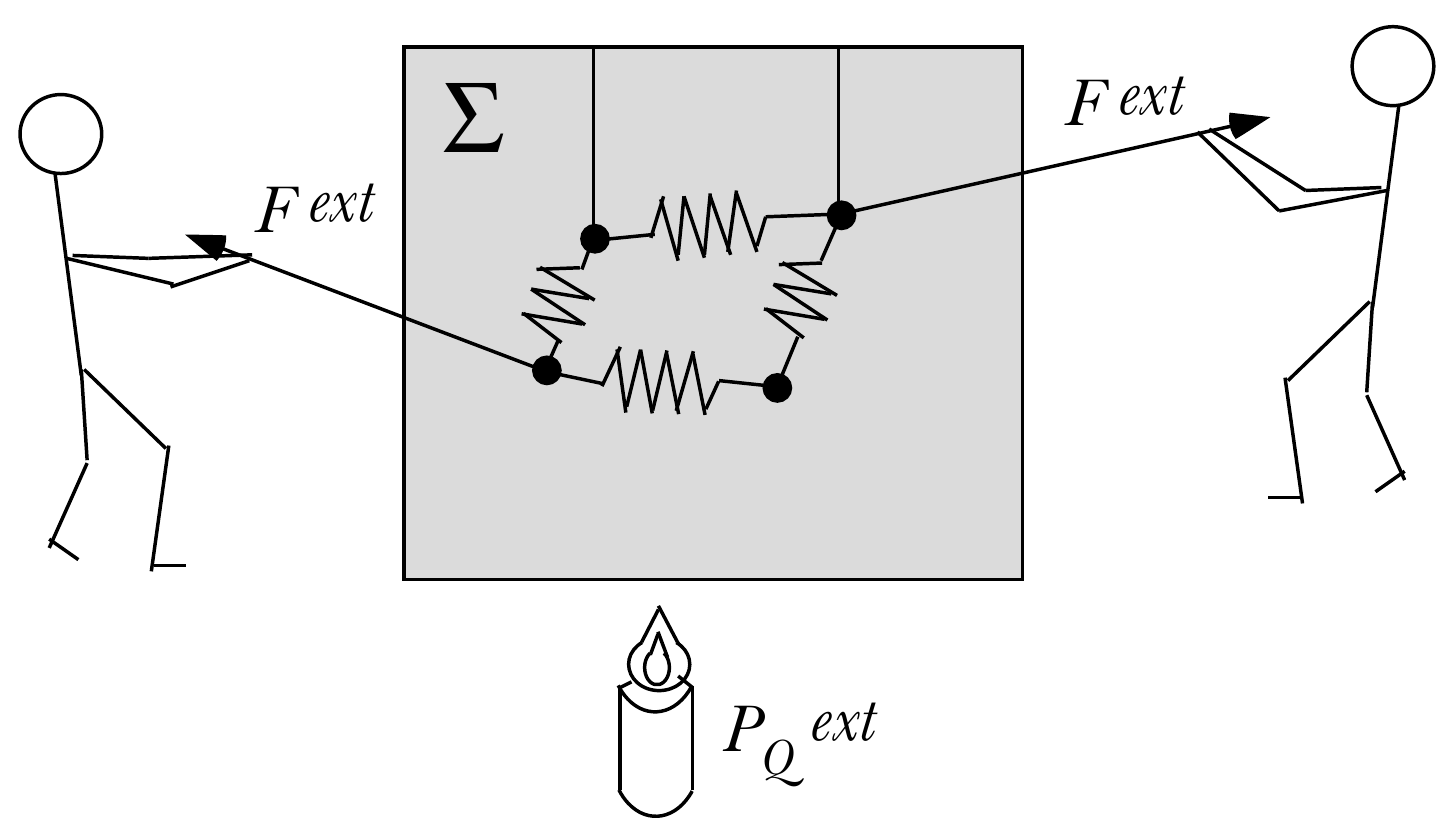}
\caption{The system $\Sigma$: $N$ particles imbedded in a
fluid.}\label{Figure}
\end{figure}

We consider a physical system $\Sigma$ of $N$ point particles
imbedded in a fluid (e.g., air, water, ...) as shown in
Figure~\ref{Figure}. The point particles are submitted to holonomic,
time-independent constraints, and we assume no a priori knowledge
for the fluid.

The system is thus defined by the particles and the fluid. It is
said to be isolated when there is no interaction between $\Sigma$
and the outside (no external force, no heat exchange, no matter
exchange). By definition, the mechanical state of $\Sigma$ is
defined by $2n$ independent variables $\xi=(\xi^1, \ldots,
\xi^{2n})=(q^1, \ldots, q^n, v^1, \ldots, v^n)$ where $q=(q^1,
\ldots, q^n)$ are the generalised coordinates and $v=(v^1, \ldots,
v^n)$, with $v^i=\frac{dq^i}{dt}$, are the generalised velocities.
Moreover, under the above condition on the system, the kinetic
energy $K$ for $N$ point particles of mass $m_{\alpha}$, defined as,
\begin{equation}
\label{en kin} K =
\frac{1}{2}\sum\limits_{\alpha=1}^{N}\,m_{\alpha}v_{\alpha}^2\ ,
\end{equation}
is expressed in terms of the generalised coordinates $q_i$ and
velocities $v_i$, by the quadratic form (see~\cite{Goldstein:2002}
Section $8.6$)
\begin{equation}
\label{quad form} K(q,v) =
\displaystyle\frac{1}{2}\sum\limits_{i,j=1}^{n}\,g_{ij}(q)\,v^iv^j
\geqslant 0\ , \qquad \mathrm{with}\quad g_{ij}=g_{ji}\ .
\end{equation}

Furthermore, the exterior of the system can act on the generalised
coordinate $q^i$ by means of a generalised force
$Q_i^{\text{\,ext}}(t)$, and the work done per unit time by this
force on $\Sigma$ is the work-power
${P_W^{\,\text{ext}}}_i(t)=Q_i^{\text{\,ext}}(t)v^i$.

From observations, we conclude that the set of mechanical variables
does not entirely define the state of the system and we are forced
to introduce thermal variables. To simplify the following
discussion, we assume that it is sufficient to add just one thermal
variable to define entirely the state. Since in the axiomatic
formulation of the second law there exists for any system a thermal
observable, the entropy $S$, we shall define the state by the set of
variables $\rho=\left(S,q,v\right)$, which are all independent by
definition of the holonomic constraints.

For our system, we assume that the energy $E$, introduced in the
first law, is the sum of the kinetic energy~\eqref{quad form} and a
potential energy $U$ independent of the velocities, {\em i.e.},
\begin{equation}
\label{split} E(S,q,v) = K(q,v) + U(S,q)\ .
\end{equation}
The ``potential energy'' $U(S,q)$ describes the internal forces but
may contain contributions from the outside of $\Sigma$, such as
potentials of conservative external forces (e.g., the gravitational
potential energy due to the earth). In this case, these forces are
considered as internal forces of $\Sigma$, and not external.

Since the system is closed, {\em i.e.}, there is no exchange of
matter with the outside, the first law of thermodynamics reduces to,
\begin{equation}
\label{first 2} \frac{dE}{dt} = P_W^{\,\text{ext}}(t) +
P_Q^{\,\text{ext}}(t)\ .
\end{equation}
Using the fact that $E=E(S,q,v)$, the LHS side of the first
law~\eqref{first 2} gives,
\begin{equation}
\label{kin 3}
\begin{split}
\frac{dE}{dt}& = \frac{\partial E}{\partial S}\,\dot{S} + \sum\limits_{i}\,\frac{\partial E}{\partial q^i}\,\dot{q}^i + \sum\limits_{i}\,\frac{\partial E}{\partial v^i}\,\dot{v}^i\\
& = \frac{\partial U}{\partial S}\,\dot{S} +
\frac{1}{2}\sum\limits_{i,j,k}\,\frac{\partial g_{ij}}{\partial
q^k}\,v^i v^j v^k + \sum\limits_{i}\,\frac{\partial U}{\partial
q^i}\,v^i + \sum\limits_{i,j}\,g_{ij}v^i\,\dot{v}^j \ .
\end{split}
\end{equation}
At this point, we must insist on the fact that the choice of general
coordinates was completely arbitrary. Therefore, we want to impose
the covariance of the time evolution equations, {\em i.e.}, they
must have the same structure for any coordinate transformation of
the mechanical variables,
\begin{align}\label{covariance}
\begin{split}
&q'^{\,i}=q'^{\,i}(q)\ ,\\
&v'^{\,i}=\sum\limits_{j}\frac{\partial q'^{\,i}}{\partial q^j}v^j\
.
\end{split}
\end{align}
Under the covariance requirement, $g_{ij}$ must be a second order,
symmetrical, covariant tensor so that $K$ is a scalar. In order to
ensure that $K$ is positive definite, $g_{ij}$ must have a positive
definite signature. Moreover, one can easily check that
\begin{equation}
\label{kin energy} \frac{dK}{dt} =
\frac{1}{2}\sum\limits_{i,j,k}\,\frac{\partial g_{ij}}{\partial
q^k}\,v^i v^j v^k + \sum\limits_{i,j}\,g_{ij}v^i\,\dot{v}^j \ ,
\end{equation}
is not covariant. However, the second term on the RHS of~\eqref{kin
3} is totally symmetric and invariant under cyclic permutation of
indices, so that we have,
\begin{equation}
\label{curvature term}
\frac{1}{2}\sum\limits_{i,j,k}\,\frac{\partial g_{ij}}{\partial
q^k}\,v^i v^j v^k = \sum\limits_{i,j,k}\,\left(\frac{\partial
g_{ij}}{\partial q^k} - \frac{1}{2}\frac{\partial g_{jk}}{\partial
q^i}\right)\,v^i v^j v^k =
\frac{1}{2}\sum\limits_{i,j,k}\,\left(\frac{\partial
g_{ij}}{\partial q^k} + \frac{\partial g_{ik}}{\partial q^j} -
\frac{\partial g_{jk}}{\partial q^i}\right)\,v^i v^j v^k\ .
\end{equation}
Introducing the symbols,
\begin{equation}
\label{Christoffel} \Gamma_{ijk}(q) =
\frac{1}{2}\,\left(\frac{\partial g_{ij}}{\partial q^k} +
\frac{\partial g_{ik}}{\partial q^j} - \frac{\partial
g_{jk}}{\partial q^i}\right)\ ,
\end{equation}
the LHS of the first law~\eqref{first 2} reduces to the covariant
equation,
\begin{equation}
\label{kin A} \frac{dE}{dt} = \frac{\partial U}{\partial S}\,\dot{S}
+
\sum\limits_{i}v^i\left(\sum\limits_{j}\,g_{ij}\,\dot{v}^j+\sum\limits_{j,k}\,\Gamma_{ijk}\,v^j
v^k + \frac{\partial U}{\partial q_i}\right)\ .
\end{equation}
As for the RHS of~\eqref{first 2}, we have~\cite{Goldstein:2002},
\begin{equation}\label{power 1}
P_W^{\,\text{ext}}(t)=\sum\limits_{i}\,Q_i^{\text{\,ext}}(t)v^i(t)\
,
\end{equation}
where $Q_i^{\,\text{ext}}(t)$ is the external generalised force
associated with $q^i$.

It is useful to introduce two new state functions, defined
respectively as,
\begin{eqnarray}
Q_i^{\,\text{int}}(S,q) &=& -\frac{\partial U}{\partial q^i}\ ,\\
T(S,q,v) &=& \frac{\partial U}{\partial S}\ ,
\end{eqnarray}
where $Q_i^{\,\text{int}}(S,q)$ is the internal force associated
with the generalised coordinate $q^i$ and $T(S,q,v)$ is called the
``temperature''~\cite{Stueckelberg:1974}. Using these definitions,
the first law~\eqref{first 2} reduces to the thermodynamic equation,
\begin{equation}
\label{kin 5} T\frac{dS}{dt} =
\sum\limits_{i}v^i\left(Q_i^{\text{\,ext}}(t) + Q_i^{\,\text{int}} -
\sum\limits_{j}\,g_{ij}\,\dot{v}^j-\sum\limits_{j,k}\,\Gamma_{ijk}\,v^j
v^k\right) + P_Q^{\,\text{ext}}(t)\ .
\end{equation}
%


\section{System of Point Particles: Second Law}
\label{4}

We investigate now the consequence of the second law for our system
by proceeding in three steps: first, we consider an isolated system,
then, we extend our analysis to an adiabatically closed system,
finally, we generalise our approach to a closed system with heat
exchange.

\subsection{Thermodynamics of an Isolated System}

An isolated system is characterised by the absence of interaction
with the exterior, {\em i.e.}, $Q_i^{\,\text{ext}}(t)=0$ and
$P_Q^{\,\text{ext}}(t)=0$. In this case, it follows from~\eqref{kin
5} that
\begin{equation}
\label{kin 5 bis} \dot{S}(t) = \frac{1}{T}\,
\sum\limits_{i}v^i\left(Q_i^{\,\text{int}} -
\sum\limits_{j}\,g_{ij}\,\dot{v}^j-\sum\limits_{j,k}\,\Gamma_{ijk}\,v^j
v^k\right)\ .
\end{equation}
This means that there exists a state function $I(S,q,v)$ called
``entropy production'' given by
\begin{equation}
\label{kin 5 tri} I(S,q,v) = \frac{1}{T(S,q,v)}\,
\sum\limits_{i}v^i\left(Q_i^{\,\text{int}} -
\sum\limits_{j}\,g_{ij}\,\dot{v}^j-\sum\limits_{j,k}\,\Gamma_{ijk}\,v^j
v^k\right)\ ,
\end{equation}
and such that
\begin{equation}
\label{state funct I} \dot{S}(t)=I(S(t),q(t),v(t))\ .
\end{equation}

At this point, we introduce the state function
$Q_i^{\,\text{fr}}(S,q,v)$, called ``friction force associated with
the generalised coordinate $q^i$'', defined as,
\begin{equation}\label{friction def}
Q_i^{\,\text{fr}} = - Q_i^{\,\text{int}} +
\sum\limits_{i}\,g_{ij}\,\dot{v}^j +
\sum\limits_{j,k}\,\Gamma_{ijk}\,v^j v^k\ .
\end{equation}
With this definition~\eqref{friction def}, the entropy
production~\eqref{kin 5 tri} can be expressed as,
\begin{equation}\label{entropy prod friction}
I(S,q,v) = -\frac{1}{T} \sum\limits_{i}v^iQ_i^{\,\text{fr}}\ .
\end{equation}
This expression of the entropy production given as the products of
the generalised velocities with the associated friction forces (or
affinities) is analogous to the usual expression in nonequilibrium
thermodynamics found for example
in~\cite{Callen:1985}~(p.~309),~\cite{deGroot:1984}~(p.~30)
or~\cite{Prigogine:1999}~(p.~213). From the evolution part of the
second law~\eqref{entropy_prod} (following the standard methods
outlined
in~\cite{Stueckelberg:1974}~p.~31,~\cite{deGroot:1984}~p.~31,~\cite{Prigogine:1999}~pp.~215,~235,~\cite{Perez:1997}~p.~348)
we must have, $I(S,q,v) \geqslant 0$, and thus the friction force is
of the form,
\begin{equation}
\label{friction coefficients} Q_i^{\,\text{fr}}(S,q,v) =
-\sum\limits_j\,\lambda_{ij}\,(S,q,v)\,v^j\ ,
\end{equation}
where the ``friction coefficient tensor $\lambda_{ij}(S,q,v)$'' must
satisfy the non-negativity condition,
\begin{equation}\label{non-negativity cond}
\frac{1}{T(S,q,v)}\,\left\{\lambda_{(ij)}(S,q,v)\right\}\geqslant 0\
,\qquad\text{where}\qquad\lambda_{(ij)}=\frac{1}{2}\left(\lambda_{ij}+\lambda_{ji}\right)\
.
\end{equation}
Note that in the usual linear scheme of non-equilibrium
thermodynamics, {\em i.e.}, near an equilibrium point, one considers
$\lambda_{ij}$ to be the coefficients of a constant matrix computed
at this equilibrium point, as explained
in~\cite{Prigogine:1999}~(p.~236).

In conclusion, in this case, the entropy production is given by,
\begin{equation}
\label{entropy quadratic} I(S,q,v) =
\frac{1}{T(S,q,v)}\sum\limits_{i,j}\,\lambda_{ij}(S,q,v)\,v^iv^j
\geqslant 0\ .
\end{equation}
Thus, we have obtained the time evolution equations for the isolated system,
\begin{equation}
\label{evo eq}
\begin{cases}
\displaystyle\sum\limits_{j}\,g_{ij}\,\ddot{q}^j + \displaystyle\sum\limits_{j,k}\,\Gamma_{ijk}\,\dot{q}^j\dot{q}^k = -\frac{\partial U}{\partial q^i} - \sum\limits_{j}\,\lambda_{ij}\dot{q}^j &\text{mechanical equations}\\
\displaystyle\frac{dS}{dt} =
\displaystyle\frac{1}{T}\sum\limits_{i,j}\,\lambda_{ij}\,\dot{q}^i\dot{q}^j
&\text{thermal equation}
\end{cases}
\end{equation}
Hence, in such a phenomenological approach, the system is characterised by the state functions
$g_{ij}(q)$, $U(S,q)$ and $\lambda_{ij}(S,q,\dot{q})$, which have to
be determined experimentally. It
should be stressed that \linebreak $Q_i^{\,\text{int}} = -\frac{\partial
U}{\partial q^i}$ is an internal conservative force (e.g., the force
exerted between two particles) introduced in the first law and that 
$Q_i^{\,\text{fr}}$ is a friction force exerted by the fluid in
which the particles are imbedded. Note that in the particular case
of solid friction, the friction force is
singular~\cite{Ferrari:2010}.

\subsection{Thermodynamics of an Adiabatically Closed System}

We now consider an adiabatically closed system, {\em i.e.},
$P_Q^{\,\text{ext}}(t)=0$, for which the thermodynamic
equation~\eqref{kin 5} reduces to,
\begin{equation}
\label{kin eq 1} T\frac{dS}{dt} =
\sum\limits_{i}v^i\left(Q_i^{\text{\,ext}}(t) + Q_i^{\,\text{int}} -
\sum\limits_{j}\,g_{ij}\,\dot{v}^j-\sum\limits_{j,k}\,\Gamma_{ijk}\,v^j
v^k\right)\ .
\end{equation}
Again, we introduce the generalised friction force,
\begin{equation}\label{friction gen adia}
Q_i^{\,\text{fr}} = - Q_i^{\,\text{ext}}(t) - Q_i^{\,\text{int}} +
\sum\limits_{j}\,g_{ij}\,\dot{v}^j +
\sum\limits_{j,k}\,\Gamma_{ijk}\,v^j v^k\ ,
\end{equation}
which yields,
\begin{equation}
T\frac{dS}{dt}= - \sum\limits_{i}v^iQ_i^{\text{\,fr}}\ ,
\end{equation}
and
\begin{equation}
\sum\limits_{j}\,g_{ij}\,\dot{v}^j +
\sum\limits_{j,k}\,\Gamma_{ijk}\,v^j v^k = Q_i^{\,\text{int}} +
Q_i^{\,\text{fr}} + Q_i^{\,\text{ext}}(t)\ .
\end{equation}
Since the system is adiabatically closed, the evolution part of the
second law~\eqref{entropy_prod} implies again,
\begin{equation}\label{entropy prod friction II}
\frac{dS}{dt} = I(t) =
-\frac{1}{T}\sum\limits_{i}v^iQ_i^{\,\text{fr}}\geqslant 0\ ,
\end{equation}
which suggests that the friction force has the same form as for the
isolated system,
\begin{equation}\label{friction isolated}
Q_i^{\,\text{fr}} = -\sum\limits_{j}\lambda_{ij}(S,q,v)v^j\ ,
\end{equation}
and thus is again a state function with the same coefficients
$\lambda_{ij}$ as in the isolated case. Similarly to the isolated
system, the time evolution equations are given by,
\begin{equation}
\label{evo eq II}
\begin{cases}
\displaystyle\sum\limits_{j}\,g_{ij}\,\ddot{q}^j + \displaystyle\sum\limits_{j,k}\,\Gamma_{ijk}\,\dot{q}^j\dot{q}^k = -\frac{\partial U}{\partial q^i} - \sum\limits_{j}\,\lambda_{ij}\dot{q}^j +  Q_i^{\,\text{ext}}(t) &\text{mechanical equations}\\
\displaystyle\frac{dS}{dt} =
\displaystyle\frac{1}{T}\sum\limits_{i,j}\,\lambda_{ij}\,\dot{q}^i\dot{q}^j
&\text{thermal equation}
\end{cases}
\end{equation}
It is worth mentioning that the arbitrary external force
$Q_i^{\,\text{ext}}(t)$ should not be confused with the internal
forces, which are the conservative force $Q_i^{\,\text{int}}$ and
the dissipative friction force $Q_i^{\,\text{fr}}$.

\subsection{Thermodynamics of a Closed System}

We now consider a closed system, {\em i.e.},
$P_{\text{chem}}^{\,\text{ext}}(t)=0$, for which the thermodynamic
equation is given by~\eqref{kin 5}.  The entropy variation of the
system is due to the friction force~\eqref{friction gen adia}, as
for the isolated and adiabatically closed systems, but also to heat
exchange with the exterior according to~\eqref{kin 5} {\em i.e.},
\begin{equation}
\label{entropy closed} T\frac{dS}{dt}= -
\sum\limits_i\,v^i\,Q_i^{\,\text{fr}} + P_Q^{\,\text{ext}}(t)\ .
\end{equation}
Hence, similarly to the isolated and adiabatically closed systems,
the time evolution equations are given~by,
\begin{equation}
\label{evo eq III}
\begin{cases}
\displaystyle\sum\limits_{j}\,g_{ij}\,\ddot{q}^j + \displaystyle\sum\limits_{j,k}\,\Gamma_{ijk}\,\dot{q}^j\dot{q}^k = -\frac{\partial U}{\partial q^i} - \sum\limits_{j}\,\lambda_{ij}\dot{q}^j + Q_i^{\,\text{ext}}(t) &\text{mechanical equations}\\
\displaystyle\frac{dS}{dt} =
\displaystyle\frac{1}{T}\sum\limits_{i,j}\,\lambda_{ij}\,\dot{q}^i\dot{q}^j
+ \displaystyle\frac{1}{T}P_Q^{\,\text{ext}}(t) &\text{thermal
equation}
\end{cases}
\end{equation}

\subsection{Equilibrium Part of the Second Law}

Finally, we can investigate the consequences of the equilibrium part
of the second law~\eqref{entropy_max}. In this case, we have to
assume that the system is isolated for all $t\geqslant t_0$ and we
have to look for the maximum of $S$ under the condition that the
energy $E(S,q,v)=\bar{E}$ is fixed. In other words, we have to look
for the maximum of $S(\bar{E},q,v)$.

One can show~\cite{Stueckelberg:1974} that this maximum condition on
$S$ is equivalent to the condition that the energy $E(S,q,v)$ is
minimum for a fixed $S=\bar{S}$ if the temperature $T$ is positive,
and is maximum if $T$ is negative, {\em i.e.},
\begin{equation}\label{ext cond}
\delta^{(1)}E\,\bigg|_{S=\bar{S}}=0\ ,\qquad\text{and}\qquad
\frac{1}{T}\,\delta^{(2)}E\,\bigg|_{S=\bar{S}}\geqslant 0\ .
\end{equation}
\begin{itemize}

\item [(i)] The extremum condition $\delta^{(1)}E\Big|_{S=\bar{S}}=0$ implies that,

\vspace{0.2cm}

\begin{itemize}

\item [$\bullet$] $\displaystyle\frac{\partial E}{\partial v^i}\bigg|_{S=\bar{S}}=0\ \,\text{{\em i.e.},}\ \,\sum\limits_{j}g_{ij}(q)v^j=0\ \,\text{and thus}\ \,v^j=0\ \text{since} \ \,g_{ij}\ \,\text{is positive definite.}$

\item [$\bullet$] $\displaystyle\frac{\partial E}{\partial q^i}\bigg|_{S=\bar{S}}=0\ \,\text{which gives (with } v^j=0\text{)}\ ,\ \,Q^{\,\text{int}}=-\frac{\partial U}{\partial q^i}=0\ .$

\vspace{0.3cm}

In other words, the extremum condition implies that the system
evolves to an equilibrium point of the time evolution
equation~\eqref{evo eq}, characterised by $q=\bar{q}$ and
$v=\bar{v}=0$.

\vspace{0.2cm}

\end{itemize}

\item [(ii)] The condition $\frac{1}{T}\delta^{(2)}E\Big|_{S=\bar{S},\,q=\bar{q},\,v=\bar{v}=0}\geqslant 0$ implies first that the matrix $\frac{1}{T}g_{ij}(\bar{q})$ is non-negative and thus, for our mechanical system with positive kinetic energy, the temperature is necessarily positive. This in turn implies that the friction coefficients matrix is non-negative. Moreover, this same condition also implies that the matrix $\frac{1}{T}\,\frac{\partial^2U}{\partial q^i\partial q^j}\left(\bar{S},\bar{q}\right)$ must be non-negative. In conclusion, the equilibrium part of the second law implies that the system evolves towards a stable equilibrium point. This is the zeroth law of thermodynamics~\cite{Stueckelberg:1974}.

\end{itemize}


\section{From Thermodynamics to Mechanics}
\label{5}

The symmetric matrix $g_{ij}$ can be identified as a metric with a
positive definite signature on the configuration space. Thus, the
configuration space is a Riemannian manifold endowed with a
torsion-free Levi-Civita connection~\cite{Weinberg:1972}, where the
components,
\begin{equation}
\label{Christoffel 2} \Gamma_{ijk}(q) =
\frac{1}{2}\,\left(\frac{\partial g_{ij}}{\partial q^k} +
\frac{\partial g_{ik}}{\partial q^j} - \frac{\partial
g_{jk}}{\partial q^i}\right)\ ,
\end{equation}
are commonly referred to as the Riemann-Christoffel symbols. Such a
manifold preserves the generalised infinitesimal distances squared
$ds^2$ defined in terms of the generalised coordinates
as~\cite{Hobson:2006},
\begin{equation}
\label{distance element} ds^2 =
\sum\limits_{i,j}\,g_{ij}(q)\,dq^idq^j\ ,
\end{equation}
and satisfies the metricity condition~\cite{Brechet:2008}, which
requires the covariant derivative of the metric with respect to
every coordinate $q^i$ to vanish,
\begin{equation}
\label{covariant derivative cond} \nabla_{i}\,g_{jk} = 0\ .
\end{equation}
In the particular case of an isolated system where the friction
coefficients matrix is strictly zero, then from the second relation
in~\eqref{evo eq}, the entropy is necessarily constant. Furthermore,
if the potential energy is zero, {\em i.e.}, $U=0$, then the
evolution of the system is a geodesic in configuration space given
by,
\begin{equation}
\label{geodesics} \dot{v}^i +
\sum\limits_{j,k}\,{\Gamma^{i}}_{jk}\,v^j v^k = 0\ ,
\end{equation}
thus satisfying the 1st law of Newton.

At this point, it is convenient to introduce the Lagrangian of the
system defined as,
\begin{equation}
\label{Lagrangian} \mathcal{L}(S,q,\dot{q}) = K(q,\dot{q}) - U(S,q)
= \frac{1}{2}\,\sum\limits_{i,j}g_{ij}\dot{q}^i\dot{q}^j - U(S,q)\ .
\end{equation}
In order to recast the dynamical terms on the RHS of the
thermodynamic equation~\eqref{kin 5} in an analytical manner, we
compute the partial derivatives of the Lagrangian and their time
derivatives,
\begin{align}
\label{derivatives I}
&\dfrac{\partial\mathcal{L}}{\partial q^i} = \dfrac{1}{2}\,\displaystyle\sum\limits_{j,k}\dfrac{\partial g_{jk}}{\partial q^i}\,\dot{q}^j\dot{q}^k - \dfrac{\partial U}{\partial q^{i}}\ ,\\
\label{derivatives II}
&\dfrac{d}{dt}\left(\dfrac{\partial\mathcal{L}}{\partial\dot{q}^i}\right)
= \displaystyle\sum\limits_{j}g_{ij}\ddot{q}^j +
\displaystyle\sum\limits_{j,k}\dfrac{\partial g_{ij}}{\partial
q^k}\,\dot{q}^j\dot{q}^k =
\displaystyle\sum\limits_{j}g_{ij}\ddot{q}^j +
\dfrac{1}{2}\,\displaystyle\sum\limits_{j,k}\,\left(\dfrac{\partial
g_{ij}}{\partial q^k} + \dfrac{\partial g_{ik}}{\partial
q^j}\right)\,\dot{q}^j\dot{q}^k\ .
\end{align}
From the differential relations~\eqref{derivatives I}
and~\eqref{derivatives II}, we derive the dynamic identity,
\begin{equation}
\label{Lagrange id}
\frac{d}{dt}\left(\dfrac{\partial\mathcal{L}}{\partial\dot{q}^i}\right)
- \frac{\partial\mathcal{L}}{\partial q^i}
=\sum\limits_{j}\,g_{ij}\,\ddot{q}^j +
\sum\limits_{j,k}\,\Gamma_{ijk}\,\dot{q}^j\dot{q}^k + \frac{\partial
U}{\partial q^i}\ .
\end{equation}
Using this identity, the thermodynamic equation~\eqref{evo eq III}
reduces to,
\begin{equation}\label{Lagrange coupling}
\begin{cases}
\dfrac{d}{dt}\left(\dfrac{\partial\mathcal{L}}{\partial\dot{q}^i}\right) - \dfrac{\partial\mathcal{L}}{\partial q^i} = Q^{\,\text{ext}}_i(t) - \displaystyle\sum\limits_j\,\lambda_{ij}\,\dot{q}^j &\text{Lagrange equations}\\
\displaystyle\frac{dS}{dt} =
\displaystyle\frac{1}{T}\sum\limits_{i,j}
\,\lambda_{ij}\,\dot{q}^i\dot{q}^j +
\frac{1}{T}P_Q^{\,\text{ext}}(t) &\text{Thermal equation}
\end{cases}
\end{equation}
In conclusion, if the state functions $U$ and $\lambda_{ij}$ are
independent of $S$, or equivalently, if the time evolution happens
at fixed temperature by contact with a thermal bath, the mechanical
Lagrange equations decouple from the thermal equation. This is the
usual case considered in mechanics.


\section{Thermodynamics of an Isolated System of Point Particles Interacting through a Harmonic~Potential}
\label{6}

As an application of the formalism we developed, we now consider the
thermodynamics of an isolated system where the state functions $U$
and $\lambda_{ij}$ are independent of $S$. The system consists of
identical point particles interacting through a harmonic potential
and is the simplest phenomenological model of a solid, where the
harmonic oscillators represent the phonons. As generalised
coordinates, we choose for simplicity cartesian coordinates of the
point particles and denote them $q=x$. Thus, the metric reduces to
the trivial Kronecker delta, {\em i.e.}, $g_{ij}(q)=\delta_{ij}$.
The kinetic energy $K(x,\dot{x})$ of the point particles per unit
mass and the harmonic interaction potential per unit mass $U(x)$ are
respectively given by,
\begin{align}
\label{kin en}
&K(x,\dot{x}) = \dfrac{1}{2}\,\displaystyle\sum\limits_{i,j}\,\delta_{ij}\,\dot{x}^i\dot{x}^j\ ,\\
\label{pot en} &U(x) =
\dfrac{1}{2}\,\omega^2\displaystyle\sum\limits_{i,j,k,l}\,\delta_{ij}\delta_{kl}\,\left(x^i-x^k\right)\left(x^j-x^l\right)\
,
\end{align}
where the coefficient $\omega^2$ is positive by the equilibrium
condition of the second law and $\omega$ represents the angular
frequency of the identical harmonic oscillators. The Lagrangian of
the system is given by,
\begin{equation}
\label{Lagrangian system}
\mathcal{L}(x,\dot{x})=\dfrac{1}{2}\,\displaystyle\sum\limits_{i,j}\,\delta_{ij}\,\dot{x}^i\dot{x}^j-\dfrac{1}{2}\,\omega^2\displaystyle\sum\limits_{i,j,k,l}\,\delta_{ij}\delta_{kl}\,\left(x^i-x^k\right)\left(x^j-x^l\right)\
.
\end{equation}
The partial derivatives of the Lagrangian and their time derivatives
are given by,
\begin{align}
\label{derivatives III}
&\dfrac{\partial\mathcal{L}}{\partial x^i} =-\omega^2\sum\limits_{j,k,l}\,\delta_{ij}\delta_{kl}\,\left(x^j - x^l\right)\ ,\\
\label{derivatives IV}
&\dfrac{d}{dt}\left(\dfrac{\partial\mathcal{L}}{\partial\dot{x}^i}\right)
=
\dfrac{d}{dt}\left(\displaystyle\sum\limits_{j}\delta_{ij}\dot{x}^j\right)
= \displaystyle\sum\limits_{j}\delta_{ij}\ddot{x}^j\ .
\end{align}
Finally, the system of coupled thermodynamical equations~\eqref{evo
eq} is explicitly found to be,
\begin{equation}
\label{Euler II}
\begin{cases}
\displaystyle\sum\limits_{j}\left[\delta_{ij}\ddot{x}^j + \lambda_{ij}\,\dot{x}^j + \omega^2\,\delta_{ij}\displaystyle\sum\limits_{k,l}\,\delta_{kl}\,\left(x^j - x^l\right) \right] = 0 &\text{Lagrange equations}\\
\displaystyle\frac{dS}{dt} =
\displaystyle\frac{1}{T}\sum\limits_{i,j}
\,\lambda_{ij}\,\dot{x}^i\dot{x}^j &\text{Thermal equation}
\end{cases}
\end{equation}
The physical interpretation of these evolution equations is
clear~\cite{Goldstein:2002}. The Lagrange equations are a system of
coupled damped harmonic oscillators where the damping term is due to
the action of a viscous friction force. These equations are in turn
coupled to the thermal equation through the friction force. Since
$\lambda_{ij}$ and $\omega^2$ do not depend on $S$, the Lagrange
equations can be solved independently to find $x(t)$, which in turn
will give $S(t)$ using the thermal equation.

If the friction matrix $\{\lambda_{ij}\}$ is positive, the condition
that the system evolves to a state of maximal entropy implies that
the system evolves towards the equilibrium state ($\bar{x}^i=0$,
$\dot{\bar{x}}^i=0$). For a strictly mechanical system, {\em i.e.},
if $\lambda_{ij}=0$, then $S$ is a constant; in this case the system
will oscillate around the equilibrium state according to,
\begin{equation}
\label{mech sys}
\begin{cases}
\ddot{x}^i + \omega^2\,\displaystyle\sum\limits_{j}\,\left(x^i - x^j\right) = 0 &\text{Lagrange equations}\\
\displaystyle S=\text{const} &\text{Thermal equation}
\end{cases}
\end{equation}
%
\section{Conclusions}

We followed the phenomenological approach developed by St\"uckelberg
to study a closed thermodynamic system consisting of a fixed number
of point particles. We thus obtained a system of coupled
differential equations describing the thermodynamic evolution of the
system. We observed that in this system of differential equations,
the coupling between the Lagrange equations and the thermal equation
is due to the viscous friction terms. If the phenomenological
variables $U$, $\lambda_{ij}$, do not depend on $S$, we recover the
usual Newton or Lagrange equations with friction. In the general
case where these variables do depend on $S$, {\em i.e.}, on
temperature, such a coupling shows that thermodynamics is a natural
extension of mechanics. Finally, we applied our coupled
thermodynamic equations to study the thermodynamics of an isolated
system consisting of identical point particles interacting through a
harmonic potential, which is the simplest phenomenological model of
a solid.

\section*{Acknowledgements}

The authors would like to honour the memory of Ernst Carl Gerlach
St\"uckelberg von Breidenbach who, among other great achievements,
developed a genuine dynamic theory of thermodynamics. Our special
thanks to one of the referees for his very valuable suggestions.


\bibliographystyle{mdpi}
\makeatletter
\renewcommand\@biblabel[1]{#1. }
\makeatother
\bibliography{references}




\end{document}